\documentclass{article}
	\usepackage{amsmath}
  \usepackage{amssymb}
  \usepackage{mathrsfs}
\usepackage{epsf}
\usepackage{epsfig}

\usepackage{amsmath}

\usepackage{amssymb}

\numberwithin{equation}{section}

\newcommand{\be}{\begin{equation}}
\newcommand{\ee}{\end{equation}}
\newcommand{\bea}{\begin{eqnarray}}
\newcommand{\eea}{\end{eqnarray}}
\newcommand{\bg}{\begin{gather}}
\newcommand{\eg}{\end{gather}}

\newcommand{\bb}{\bibitem}
 
\newcommand{\eqn}{\begin{eqnarray}}
\newcommand{\eqnx}{\end{eqnarray}}

\begin{document}

\author{{\L}. T. St\c{e}pie\'{n}\footnote{The Pedagogical University of Cracow, ul. Podchorazych 2, 30-084 Krakow, Poland, e-mail: sfstepie@cyf-kr.edu.pl}}

\date{}

\title{On Bogomolny equations in generalized gauged baby BPS Skyrme models}

\maketitle 

\begin{abstract}
Using the  concept of strong necessary conditions (CSNC), we derive Bogomolny equations and BPS bounds 
for two modifications of the gauged baby BPS Skyrme model: the nonminimal coupling to the gauge field and k-deformed 
model. In particular, we study, how the Bogomolny equations and the equation for the potential, reflect these two modifications. In both examples, the CSNC method shows to be a very useful tool. 
\end{abstract}

\section{Introduction}
BPS models i.e., field theories which allow for solitonic solutions and simultaneously admit a reduction of the full second order static equation of motion to a set of first order equations (Bogomolny or BPS equations \cite{BPS}-\cite{PraSom}), play a notable role in current physics. The appearance of Bogomolny equations not only leads to exact solutions, which significantly enlarge our understanding of considered non-linear models, but also guarantee existence of a topological Bogomolny bound, which results in a topological stability of solitons carrying a non-trivial value of the corresponding topological charge. Therefore, it is of high importance to search for models with the BPS property. In fact, between few known rather restricted methods (completing to a square \cite{BPS}-\cite{Bog gen}, the first order formalism \cite{bazeia}, on-shell method \cite{onshell}), there exists a completely general method, which allows for a systematic derivation (if possible) BPS equations. This method, referred as  {\it the concept of strong necessary conditions} (CSNC),  was originally proposed and analyzed in \cite{sokalski1979}-\cite{sokalski2009}, has been very recently further developed by Adam and Santamaria \cite{adam}, who proposed so called {\it first order Euler-Lagrange} (FOEL) formalism.  

In the present work we want to apply the CSNC method to a generalized (gauged) baby BPS Skyrme model. The baby BPS Skyrme model \cite{BPS Zak}-\cite{stepien2012} is a limit of the full baby Skyrme theory \cite{old}, where the usual 2+1 dimensional Dirichlet term $\mathcal{L}_2= \partial_\mu \vec{\phi} \cdot \partial^\mu \vec{\phi} $ disappears. (The full baby Skyrme  model \cite{MantonSutcliffe}, is a planar analogon of the Skyrme model, \cite{skyrme}, \cite{GP3}, \cite{MakhankovEtal}, \cite{MantonSutcliffe}). Here the iso-vector field $\vec{\phi}$ is an element of two-dimensional target space $\mathbb{S}^2$. The resulting Lagrange density consists of two remaining parts: the $2+1$ dimensional Skyrme term $\mathcal{L}_4$ and a non-derivative term i.e., a potential $V$
\be
\mathcal{L}_{BPS\; bSk} = \alpha_4 \mathcal{L}_4+\alpha_0 \mathcal{L}_0\equiv - \alpha_4 
\left( \partial_\mu \vec{\phi} \times \partial_\nu \vec{\phi}  \right)^2 - \alpha_0V(\vec{n}\cdot \vec{\phi})
\ee
This model possesses a BPS equation which is saturated by infinitely many topological solitons (baby Skyrmions) carrying arbitrary value of the pertinent topological charge $Q \in \pi_2(\mathbb{S}^2)$ \cite{BPS baby}, \cite{stepien2012}, \cite{Speight}. In fact, this equation is a two dimensional analogous of the famous BPS equation for the BPS Skyrme model \cite{Skyrme} (see also \cite{stepien2016}-\cite{bjarke}), which plays an important role in a possible solution of the too large binding energy problem in Skyrme model \cite{binding}. The next obvious step in analysis of the baby BPS Skyrme model was the minimal coupling of it to the Maxwell $U(1)$ gauge field, and study of magnetic properties of resulting gauged baby BPS Skyrme model \cite{stepien2015}, \cite{gBPS baby} 
\be
\mathcal{L}_{g \;BPS\; bSk} = - \alpha_4 
\left( D_\mu \vec{\phi} \times D_\nu \vec{\phi}  \right)^2 - \alpha_0V(\vec{n}\cdot \vec{\phi}) - \alpha_{m} F_{\mu \nu}^2
\ee
where the usual derivatives are change into the covariant ones $D_\mu \vec{\phi} = \partial_\mu \vec{\phi}+A_\mu \vec{n} \times \vec{\phi} $ as there is unbroken $U(1)$ subgroup of the $SO(3)$ iso-rotations of $\vec{\phi}$. It has been shows that this model also revels the BPS property where a BPS equation requires a non-trivial {\it superpotential} $W$ which is related to the original potential by a target space equation   
\be
\lambda^2 W'^2 + g^2  \lambda^4 W^2 =2\mu^2 V
\ee
where the prime denotes differentiation w.r.t. $\phi_3$. (Here $\alpha_4=\frac{\lambda^2}{4}, \alpha_0=\mu^2, \alpha_m=\frac{1}{4g^2}$.) In a consequence, the gauging of the model provided some restriction on the potential as far as one wants to keep the BPS property. For example, the BPS equation cannot be satisfied, if a two vacua potential is assumed. This is in a contrast to the non-gauged case, where all (reasonable) potentials lead to saturated solutions of the BPS equation i.e., baby BPS Skyrmions. 

Here we want to further analyze, how the existence and the form of the BPS equation (and superpotential) is affected by 1) non-minimal coupling to the gauge field and/or by 2) k-extension of the model i..e, if the baby Skyrme 4-derivate term is replaced by a function of it. Both modifications are motivated by effective nature of the baby Skyrme model. In such a case, non-standard kinetic part (k-model generalization) as well as addition of a "dielectric" function to the gauge field part are typically accepted generalizations - see for example, condense matter \cite{k-baby} or cosmological applications \cite{k}. Obviously, the lagrangian should be gauge invariant, however in the case of gauged models investigated in this paper, we want also to investigate (as in \cite{stepien2015}), whether any terms analogical to Proca terms (\cite{arodzhadasz}, \cite{rubakov}), will appear in the expressions for the potentials. 

\section{Non-minimal extension of the gauged baby BPS Skyrme model}

\subsection{A short introduction}

The static energy functional of the non-minimally extended model is
\begin{eqnarray}
E&=&\frac{1}{2} E_0 \int d^2 x \left[ \lambda^2 \left( D_{1} \vec{\phi} \times D_{2} \vec{\phi} \right)^2 +2 \mu^2 V(\vec{\phi} \cdot \vec n) + \frac{1}{g^2} U(\vec{\phi} \cdot \vec{n}) B^2 \right]  \label{model1}\\
&=& \frac{1}{2} E_0 \int d^2 x \left[ \lambda^2 Q^2 +2 \mu^2 V(\phi_3) + \frac{1}{g^2} U(\phi_3) B^2 \right]
\end{eqnarray}
where $ B \equiv F_{12} =  \partial_{1} A_{2} - \partial_{2} A_{1}$ and 

\begin{equation}
Q\equiv \vec \phi \cdot D_1 \vec \phi \times D_2 \vec \phi=q + \epsilon_{ij} A_i \partial_j (\vec n \cdot \vec \phi), 
\end{equation}

where

\begin{equation}
q \equiv \vec \phi \cdot \partial_1 \vec \phi \times \partial_2 \vec \phi
\end{equation}

is the topological density multiplied by 4. 

\subsubsection{The case with axially symmetric ansatz}

Now we apply the CSNC method.
At the beginning, for simplicity reasons, we compute the Bogomolny equations in an axially symmetric ansatz, for any dielectric coupling function $U$:
\begin{equation} \label{rad-ans}
\vec{\phi} (r,\phi)  = \left( 
\begin{array}{c}
\sin f(r) \cos k\phi \\
\sin f(r) \sin k\phi \\
\cos f(r)
\end{array}
\right), \;\;\;\; A_{0}=0, A_{r}=ka_{1}(r), \;\;\; A_{\phi}=ka_{2}(r) 
\end{equation}
Here $r, \phi$ are polar coordinates and $k\in \mathbb{Z}$. The magnetic field is $B=\frac{ka'_{2}(r)}{r}$. Then the energy functional reads (we have assumed (for  some generality, because we want to investigate, whether any terms analogical to Proca terms, will appear) that $V =V(a_{2}, h)$): 
\begin{equation}
 \tilde{H}= 2\pi E_0\int dy \left[2  \lambda^2  k^2 (1+a_{2})^2 h_y^2 +  \mu^2 V(a_{2}, h) + \frac{1}{2g^2} U(h) k^2 a^{2}_{2,y} + F_h h_y + F_{a_{2}} a_{2,y}\right],
\end{equation}

where we introduced a new base space coordinate $y=\frac{r^2}{2}$ as well as a new target spaces variable $h = \frac{1}{2}(1-\cos f)$.
In this case, the CSNC method is equivalent to adding only the divergence: $\frac{d F(a_{2}, h)}{dy}$ (so we have here only so-called, divergent invariant), to the density of the energy functional. Hence, we have:
\begin{gather}
\frac{\partial \tilde{\mathcal{H}}}{\partial h} :  \ \mu^{2} V_{,h} + \frac{1}{2g^{2}} k^{2}a^{2}_{2,y} U_{,h} + F_{,hh} h_{,y} + F_{,a_{2}h} a_{2,y} = 0,  \label{gorne1_I} \\
\frac{\partial \tilde{\mathcal{H}}}{\partial a_{2}} : \ \mu^{2} V_{,a_{2}} + 4\lambda^{2} k^{2}(1+a) h^{2}_{,y} + F_{,a_{2}a_{2}} a_{2,y} + F_{,ha_{2}} h_{,y}= 0, \label{gorne2_I} \\
\frac{\partial \tilde{\mathcal{H}}}{\partial h_{,y}} : \ 4\lambda^{2} k^{2} (1+a_{2})^{2} h_{,y} + F_{,h} = 0,  \label{dolne1_I} \\
\frac{\partial \tilde{\mathcal{H}}}{\partial a_{2,y}} : \ \frac{k^{2}}{g^{2}}U a_{2,y} + F_{,a_{2}}=0  \label{dolne2_I}
\end{gather}

 By using the relations

\begin{gather}
h_{,y} = -\frac{F_{,h}}{4 k^{2} \lambda^{2} (1+a)^{2}}, \label{rBgm1_ansatz}\\
a_{2,y} = -\frac{g^{2} F_{,a_{2}}}{k^{2}U(h)}, \label{rBgm2_ansatz}
\end{gather}

obtained from (\ref{dolne1_I}) - (\ref{dolne2_I}), we eliminate the derivatives of $h$ and $a$ from (\ref{gorne1_I}) - (\ref{gorne2_I}). Hence, we have a system for $U, V, F$. 
In the case: $U=1$, the solution is

\begin{gather}
V(a_{2},h) = \frac{4 g^{2} \lambda^{2} (1+a_{2})^{2} (F_{,a_{2}} )^{2} + (F'_{,h})^{2} + 8c_{1} \mu^{2} k^{2} \lambda^{2} (1+a_{2})^{2}}{8 \mu^{2} k^{2} \lambda^{2} (1+a_{2})^{2}},
\end{gather}

where $F=F(a_{2},h) \in \mathcal{C}^{2}$. Of course, the dependance of the potential $V$ on $a_{2}$, disappears, when $F(a_{2},h) = c_{0} (1 + a_{2}) W(h), c_{0} = const$, so we have the equation with the superpotential $W$:

\begin{equation}
V = \frac{1}{8 \mu^{2} k^{2} \lambda^{2}} [4g^{2} \lambda^{2} c_{0} W(h) + c^{2}_{0} W'(h) ] \label{bez_U}
\end{equation}

If $U = U(h)$, we obtain two sets of the solutions:

\begin{gather}
F(a_{2},h) = F_{1}(h), \\
U=U(h) \in \mathcal{C}^{2}, \\
V(a,h) = \frac{(F'_{1,h})^{2}}{8\mu^{2} k^{2} \lambda^{2} (1+a_{2})^{2}}
\end{gather}

and 

\begin{gather}
F(a_{2},h) = f_{1}(h) a_{2} + f_{2}(h), \ U(h) = -\frac{4 f^{2}_{1}(h) g^{2} \lambda^{2}}{(f'_{1}(h))^{2} - 8  f_{3}(h) k^{2} \lambda^{2} \mu^{2} - 4 c_{3} g^{2} \lambda^{2}}, \label{z_U1} \\
V(a_{2}, h) = \frac{(f'_{1}(h) - f'_{2}(h)) \bigg(\frac{f'_{1}(h) - f'_{2}(h)}{2(1+a_{2})^{2}} - \frac{f'_{1}(h)}{1 + a_{2}} \bigg)}{4\mu^{2} k^{2} \lambda^{2}} + f_{3}(h),
\end{gather}

where $f_{k} = f_{k}(h) \in \mathcal{C}^{2} (k =1, 2, 3)$ are arbitrary functions of $h$. By eliminating of $f_{3}$ from $V$, we get:

\begin{equation}
\begin{gathered}
V(a_{2}, h) = \frac{(f'_{1}(h) - f'_{2}(h)) \bigg(\frac{f'_{1}(h) - f'_{2}(h)}{2(1+a_{2})^{2}} - \frac{f'_{1}(h)}{1 + a_{2}} \bigg)}{4\mu^{2} k^{2} \lambda^{2}} + \label{z_U2} \\ 
\frac{g^{2} f^{2}_{1}}{2\mu^{2} k^{2} U} + \frac{1}{8\mu^{2} k^{2} \lambda^{2}} (f'_{1})^{2} - \frac{c_{0} g^{2}}{2 \mu^{2} k^{2}}.
\end{gathered} 
\end{equation}

 Obviously, if $f_{1} = f_{2}$, then in the case of (\ref{z_U2}), the superpotential equation has the form  (when $W(h) \equiv f_{2}$): 

\begin{gather}
V(h) = \frac{g^{2} W^{2}}{2\mu^{2} k^{2} U} + \frac{1}{8\mu^{2} k^{2} \lambda^{2}} (W')^{2} - \frac{c_{00} g^{2}}{2 \mu^{2} k^{2}}, c_{00} = const, \label{z_U22}
\end{gather}

which is some generalization of (\ref{bez_U}).

Hence, the equations (\ref{rBgm1_ansatz}) - (\ref{rBgm2_ansatz}), are the Bogomolny decomposition for this case.

 \subsubsection{The case with stereographic variables} 

 Now we apply the stereographic projection, for the functional (\ref{model1})
 
 \begin{equation}
 \vec{\phi} = \bigg[\frac{\omega + \omega^{\ast}}{1+\omega \omega^{\ast}}, \frac{-i \cdot (\omega-\omega^{\ast})}{1+\omega \omega^{\ast}}, \frac{1 - \omega 
 \omega^{\ast}}{1+\omega \omega^{\ast}}\bigg], \ \ i.e. \ \ \omega = \frac{\phi_{1}+i \phi_{2}}{1+\phi_{3}},
 \label{stereograf}
 \end{equation} 
 
 where $\omega=\omega(x,y) \in \mathbb{C}$, $x, y \in \mathbb{R}$.\\ 

 Hence, after some rescaling, ({\em{cf}} \cite{stepien2015}) 
 
  \begin{equation}
  \begin{gathered}
   E = \int \mathcal{H} d^{2}x=\int \bigg\{ 4\lambda_{1} \frac{[i \cdot (\varepsilon^{jk}\omega_{,j}\omega^{\ast}_{,k}) - \varepsilon^{pr} A_{p} \cdot    
  (\omega \omega^{\ast})_{,r}]^{2}}{(1+\omega \omega^{\ast})^{4}} + \lambda_{2}  U B^{2} + V \bigg\} d^{2}x,
  \end{gathered}
  \end{equation}


  where $B$ is magnetic field and $B \equiv F_{12} = \partial_{1} A_{2} - \partial_{2} A_{1}$.  We have assumed here that $V =V(\omega, \omega^{\ast}, A_{k}), (k = 1, 2)$, and $U=U(\omega, \omega^{\ast})$.

 We make the following gauge transformation of $\mathcal{H}$, on the sum of the invariants $\sum^{3}_{n=1} I_{n}$, \cite{stepien2015}
  
  \begin{equation}
  \begin{gathered}
  \mathcal{H} \longrightarrow \tilde{\mathcal{H}}= \mathcal{H} +  \sum^{3}_{n=1} I_{n} = 4\lambda_{1} \frac{[i \cdot  
  (\varepsilon^{jk}\omega_{,j}\omega^{\ast}_{,k}) - \varepsilon^{pr} A_{p} \cdot  ( \omega \omega^{\ast})_{,r}]^{2}}{(1+\omega  
  \omega^{\ast})^{4}} + \lambda_{2}  U B^{2} + V + \\ 
  \lambda_{3} \{G'_{1} [(i\varepsilon^{jk}\omega_{,j}\omega^{\ast}_{,k}) - \varepsilon^{pr} A_{p} \cdot  (\omega \omega^{\ast})_{,r}] + G_{1} B \} + \sum^{2}_{l=1}  D_{l} G_{l+1}, \label{przecech}
  \end{gathered}
  \end{equation}

  where $I_{1}$ is given by: $ I_{1} = \lambda_{3} \{G'_{1} [(i\varepsilon^{jk}\omega_{,j}\omega^{\ast}_{,k}) - \varepsilon^{pr} A_{p} \cdot  ( \omega \omega^{\ast})_{,r}] + G_{1} B \}$, $\lambda_{3}=const$., $G_{1}=G_{1}(\omega \omega^{\ast}) \in \mathbb{R}$ is some arbitrary function differentiable at least twice. $G'_{1}$ denotes the derivative of the function $G_{1}$ with respect to its argument: $\omega\omega^{\ast}$, and $I_{2}= D_{x} G_{2}(\omega, \omega^{\ast}, A_{1}, A_{2}), I_{3}=D_{y} G_{3}(\omega, \omega^{\ast}, A_{1}, A_{2})$. Next: $x_{1}=x, x_{2}=y, D_{x} \equiv \frac{d}{dx}, D_{y} \equiv \frac{d}{dy}$,  and $G_{l+1}=G_{l+1}(\omega, \omega^{\ast}, A_{1}, A_{2})$, ($l = 1, 2$), are some functions (differentiable at least twice), which are to be determinated later.

After applying the concept of strong necessary conditions to (\ref{przecech}), we obtain the so-called dual 
  equations ({\it cf.} \cite{stepien2015}):
  
  \begin{eqnarray}
  \begin{gathered}
  \tilde{\mathcal{H}}_{,\omega} : 
  -16 \lambda_{1} \frac{[i \cdot (\varepsilon^{jk}\omega_{,j}\omega^{\ast}_{,k}) - 
 \varepsilon^{pr} A_{p} \cdot  ( \omega \omega^{\ast})_{,r}]^{2}}{(1+\omega 
  \omega^{\ast})^{5}} \omega^{\ast}  - N_{3}  \varepsilon^{pr} A_{p} \omega^{\ast}_{,r} +  U_{,\omega} B^{2} + V_{,\omega} +  \\
  \lambda_{3}\bigg\{G''_{1} \omega^{\ast}  
  [i \cdot (\varepsilon^{jk}\omega_{,j}\omega^{\ast}_{,k}) - 
 \varepsilon^{pr} A_{p} \cdot  (\omega \omega^{\ast})_{,r}] - \label{gorne1}    
  G'_{1}  \varepsilon^{pr} A_{p} \omega^{\ast}_{,r} + 
  G'_{1} \omega^{\ast} B \bigg\} + \\ 
  \sum^{2}_{l} D_{l}G_{l+1,\omega} = 0,
   \end{gathered} 
   \end{eqnarray}
  
    \begin{eqnarray}
  \begin{gathered}
  \tilde{\mathcal{H}}_{,\omega^{\ast}} : 
  -16 \lambda_{1} \frac{[i \cdot (\varepsilon^{jk}\omega_{,j}\omega^{\ast}_{,k}) - 
 \varepsilon^{pr} A_{p} \cdot  ( \omega \omega^{\ast})_{,r}]^{2}}{(1+\omega 
  \omega^{\ast})^{5}} \omega  - N_{3}  \varepsilon^{pr} A_{p} \omega_{,r} + U_{,\omega^{\ast}} B ^{2} + V_{,\omega^{\ast}} +  \\ 
\lambda_{3}\bigg\{G''_{1} \omega  
  [i \cdot (\varepsilon^{jk}\omega_{,j}\omega^{\ast}_{,k}) - 
 \varepsilon^{pr} A_{p} \cdot  ( \omega \omega^{\ast})_{,r}] - \label{gorne2}    
  G'_{1}  \varepsilon^{pr} A_{p} \omega_{,r} + 
  G'_{1} \omega B \bigg\} + \\ 
   \sum^{2}_{l} D_{l}G_{l+1,\omega^{\ast}} =0,
   \end{gathered} 
   \end{eqnarray}

    \begin{equation}
  \begin{gathered}
  \tilde{\mathcal{H}}_{,A_{s}}  : \
   -N_{3} (\varepsilon^{sr}(\omega \omega^{\ast})_{,r})  + V_{,A_{s}} -  \lambda_{3}G'_{1}(\varepsilon^{sr} (\omega \omega^{\ast})_{,r}) +  \sum^{2}_{l} D_{l}G_{l+1,A_{s}} = 0, 
  \label{gorne3}
  \end{gathered}
  \end{equation}

	\vspace{0.1 in}
	
  \begin{equation}
  \begin{gathered}
  \tilde{\mathcal{H}}_{,\omega_{,s}}  : \
   N_{3} \ ( i \varepsilon^{sk} \omega^{\ast}_{,k} - \varepsilon^{ps} A_{p}\omega^{\ast}) + 
  \lambda_{3}G'_{1} ( i \varepsilon^{sk} \omega^{\ast}_{,k} - \varepsilon^{ps} A_{p}\omega^{\ast}) + G_{s+1,\omega} = 0, \label{dolne_1} 
  \end{gathered}
  \end{equation}

  \begin{equation}
  \begin{gathered}
  \tilde{\mathcal{H}}_{,\omega^{\ast}_{,s}}  : \
   N_{3}  \  (i\varepsilon^{js}\omega_{,j} - \varepsilon^{ps} A_{p}\omega) + 
  \lambda_{3}G'_{1} \ (i\varepsilon^{js}\omega_{,j} - \varepsilon^{ps} A_{p}\omega) + G_{s+1,\omega^{\ast}} = 0, \label{dolne3} 
  \end{gathered}
  \end{equation}

  \begin{equation}
  \begin{gathered}
  \tilde{\mathcal{H}}_{,A_{s,r}}  : \  2\lambda_{2} U B \varepsilon^{rs} + \lambda_{3} G_{1} \varepsilon^{rs} + G_{r+1,A_{s}} = 0,
  \label{dolne5555}
  \end{gathered}
  \end{equation}

  where $N_{3} =  \frac{8\lambda_{1}[i \cdot (\varepsilon^{jk}\omega_{,j}\omega^{\ast}_{,k}) - 
 \varepsilon^{pr} A_{p} \cdot  (\omega \omega^{\ast})_{,r}]}{(1+\omega\omega^{\ast})^{4}}$ and $G'_{1}, G''_{1}$ denote the derivatives of the function $G_{1}$ with respect to its argument: $\omega\omega^{\ast}$.

   Now, we consider $\omega, \omega^{\ast}, A_{i}, (i=1,2), G_{k}$, ($k=1, 2, 3$), as equivalent dependent variables, governed by the system of equations
  (\ref{gorne1}) - (\ref{dolne5555}). We make two operations (similar operations were made firstly in \cite{stepien2015}, in the cases of gauged baby Skyrme models: full and restricted one). 
  
  Namely, as we see, after putting ({\it cf.} \cite{stepien2015}):
  
  \begin{gather}
  G'_{1} = -\frac{8\lambda_{1} [i \cdot (\varepsilon^{jk}\omega_{,j}\omega^{\ast}_{,k}) - 
 \varepsilon^{pr} A_{p} \cdot  ( \omega \omega^{\ast})_{,r}]}{\lambda_{3}(1+\omega\omega^{\ast})^{4}}, \label{warG1} \\
     B = -\frac{1}{2\lambda_{2} U} (\lambda_{3} G_{1}+G_{2,A_{2}}), \label{A2xA1y} \\ 
  G_{3,A_{1}} = - G_{2,A_{2}},  \ \  G_{2}= c_{2}A_{2}, \hspace{0.08 in} G_{3}=-c_{2}A_{1}, \ c_{2}=const, \label{warG2G3}
  \end{gather}

  the equations (\ref{dolne_1}) - (\ref{dolne5555}) become the tautologies and the candidate for Bogomolny decomposition is ({\it cf.} \cite{stepien2015}):
  
  \begin{equation}
  \begin{gathered}
  \frac{8\lambda_{1}[i \cdot (\varepsilon^{jk}\omega_{,j}\omega^{\ast}_{,k}) - 
 \varepsilon^{pr} A_{p} \cdot  ( \omega \omega^{\ast})_{,r}]}{\lambda_{3}(1+\omega\omega^{\ast})^{4}} = -G'_{1}, \label{dekBogom_restr}\\
  2\lambda_{2} U B + \lambda_{3} G_{1}+c_{2}= 0.
  \end{gathered}
  \end{equation}

    Now, the next step is checking, when the equations (\ref{gorne1}) - (\ref{gorne3}) are satisfied, if (\ref{dekBogom_restr}) hold. Thus, we insert  (\ref{warG2G3}) and (\ref{dekBogom_restr}), 
  into (\ref{gorne1}) - (\ref{gorne3}). Hence, we get some system of partial differential equations for $V$. It has turned out that $V=V(\omega, \omega^{\ast})$, and the solution of this system, for $U=U(\omega\omega^{\ast})$,  (cf. \cite{stepien2015}), is: 

    \begin{equation}
    \begin{gathered}
  V(\omega, \omega^{\ast}) = \frac{(c_{2} + \lambda_{3} G_{1}(\omega \omega^{\ast}))^{2}}{4 \lambda_{2} U(\omega \omega^{\ast})} + \\  
  \frac{\lambda^{2}_{3}}{8\lambda_{1}} \int ((1+\omega \omega^{\ast})^{3} G'_{1} \omega^{\ast} (2G'_{1} + G''_{1} (1 + \omega \omega^{\ast}))) d\omega - \\
  \frac{\lambda^{2}_{3}}{8 \lambda_{1}} \int \int \{ [ (2 + 8 \omega \omega^{\ast}) (G'_{1})^{2} + ((1 + 9 \omega \omega^{\ast}) G''_{1} + 
   G'''_{1} \omega \omega^{\ast} (1 + \omega \omega^{\ast})) (1 + \omega \omega^{\ast}) G'_{1} + \\
 \omega \omega^{\ast} (1 + \omega \omega^{\ast})^{2} (G''_{1})^{2} (1 + \omega \omega^{\ast})^{2} d \omega ] + (1+\omega \omega^{\ast})^{3} G'_{1} \omega (2G'_{1} + G''_{1} (1 + \omega \omega^{\ast})) \} \ d \omega^{\ast}.
  \end{gathered}
  \end{equation}



\section{k-deformation}
Another generalization of the gauged baby BPS Skyrme model is given by the following energy integral
\begin{equation}
E= \frac{1}{2} E_0 \int d^2 x \left[ \lambda^2 F(Q^2) +2 \mu^2 V(\phi_3) + \frac{1}{g^2} B^{2} \right]
\end{equation}

where we have again the magnetic field $B \equiv F_{12} =  \partial_{1} A_{2} - \partial_{2} A_{1}$ and the standard $Q^2$ derivative part is replaced by an arbitrary, positive definite function $G_{0}$. We call it as k-generalized gauged baby BPS Skyrme model. In fact, k-deformed field theories have been extensively investigated especially in the context of topological solitons: kinks in 1+1 dimensions \cite{k-model}-\cite{k-brane}, vortices \cite{vortex} and monopoles \cite{monopole}.


\subsection{The non-gauged version}

The next step is to consider the non-gauged version of the k-generalized model that is the k-generalized baby BPS Skyrme model
\begin{equation}
E= \frac{1}{2} E_0 \int d^2 x \left[ \lambda^2 G_{0}(q^2) +2 \mu^2 V(\phi_3)  \right]
\end{equation}
where again we restrict ourselves to the monomial function only $G_{0}(q^2)=(q^2)^n$.

After making stereographic projection, we have

\begin{equation}
E = \int \mathcal{H} d^{2} x = \int \bigg[ \bigg(4 \lambda_{1} \frac{(\varepsilon^{jk} i \ \omega_{,j}\omega^{\ast}_{,k})^{2}}{(1+\omega\omega^{\ast})^{4}}\bigg)^{n} + V(\omega, \omega^{\ast})  \bigg] d^{2} x
\end{equation}

We make the gauge transformation and we have

 \begin{gather}
 \tilde{\mathcal{H}} = \bigg(\frac{4\lambda_{1}(\varepsilon^{jk} i \ \omega_{,j}\omega^{\ast}_{,k})^{2}}{(1+\omega\omega^{\ast})^{4}}\bigg)^{n} + 
 V(\omega, \omega^{\ast}) + i G_{1}(\omega, \omega^{\ast})  \varepsilon^{jk} \omega_{,j}\omega^{\ast}_{,k} + \sum^{2}_{l=1} D_{l} G_{l+1}(\omega, \omega^{\ast})
  \end{gather} 

The dual-equations are:

  \begin{gather}
  \tilde{\mathcal{H}}_{,\omega}: -4n\frac{(4\lambda_{1})^{n} (\varepsilon^{jk} i \ \omega_{,j}\omega^{\ast}_{,k})^{2n}}{(1+\omega\omega^{\ast})^{4n+1}} \omega^{\ast} + V_{,\omega} + i G_{1,\omega} \varepsilon^{jk} \omega_{,j}\omega^{\ast}_{,k} +   
  \sum^{2}_{l=1}  D_{l} G_{l+1,\omega}=0,
  \label{gorne_kdeform_ng1} \\
 \tilde{\mathcal{H}}_{,\omega^{\ast}}: -4n\frac{(4\lambda_{1})^{n} (\varepsilon^{jk} i \ \omega_{,j}\omega^{\ast}_{,k})^{2n}}{(1+\omega\omega^{\ast})^{4n+1}} \omega + V_{,\omega^{\ast}} + i G_{1,\omega^{\ast}} \varepsilon^{jk} \omega_{,j}\omega^{\ast}_{,k} + \sum^{2}_{l=1}  D_{l} G_{l+1,\omega^{\ast}} = 0, \label{gorne_kdeform_ng2} \\
 \tilde{\mathcal{H}}_{,\omega_{,r}}:  2n\frac{(4\lambda_{1})^{n}(\varepsilon^{jk} i \ \omega_{,j}\omega^{\ast}_{,k})^{2n-1}}{(1+\omega\omega^{\ast})^{4n}}  
 \varepsilon^{rm} i \omega^{\ast}_{,m} + i G_{1} \varepsilon^{rm} \omega^{\ast}_{,m} + G_{r+1,\omega} = 0,  \label{dolne31} \\
 \tilde{\mathcal{H}}_{,\omega^{\ast}_{,r}}: 2n\frac{(4\lambda_{1})^{n} (\varepsilon^{jk} i \ \omega_{,j}\omega^{\ast}_{,k})^{2n-1}}{(1+\omega\omega^{\ast})^{4n}}  
 \varepsilon^{mr} i \omega_{,m} + i G_{1} \varepsilon^{mr} \omega_{,m} + G_{r+1,\omega^{\ast}} = 0.  \label{dolne32}
 \end{gather} 

 The equations (\ref{dolne31}) - (\ref{dolne32}) become the tautologies, if we put:

 \begin{eqnarray}
 (\varepsilon^{jk} i \ \omega_{,j}\omega^{\ast}_{,k})^{2n-1} = -\frac{(1+\omega\omega^{\ast})^{4n} G_{1}}{2n(4 \lambda)^{n}}, \label{uzg1} \\
 G_{r+1} = const, \ r = 1, 2  \label{uzg2}
 \end{eqnarray} 
 
 Next, we eliminate the derivatives of the fields $\omega_{,k}, \omega^{\ast}_{,k} \ (k = 1, 2)$, from (\ref{gorne_kdeform_ng1}) - (\ref{gorne_kdeform_ng2}), by using (\ref{uzg1}) - (\ref{uzg2}). We get the system of equations for $V(\omega, \omega^{\ast})$. We find the solution of this system, for $G_{1}=G_{1}(\omega \omega^{\ast} )$: 

 \begin{equation}
 \begin{gathered}
 V(\omega, \omega^{\ast}) = \int \bigg((1+\omega \omega^{\ast})^{-1} \bigg( \omega^{\ast} \bigg( -2^{2(n+1)} n  \bigg(-\frac{\lambda_{1}(\omega \omega^{\ast} + 1)^{4n} G_{1}}{2n(4\lambda_{1})^{n}}\bigg)^{\frac{2n}{2n-1}}\\ 
(\omega \omega^{\ast} + 1)^{-4n} +  \bigg(-\frac{\lambda_{1}(\omega \omega^{\ast} + 1)^{4n} G_{1}}{2n(4\lambda_{1})^{n}}\bigg)^{\frac{1}{2n-1}} G'_{1} (1 + \omega \omega^{\ast}) \bigg) \bigg) \bigg) d\omega + \\
\int \frac{1}{(2n-1) (\omega \omega^{\ast} + 1)} \bigg\{ (\omega \omega^{\ast} + 1) \bigg[ \int \frac{1}{G_{1} (\omega \omega^{\ast} + 1)^{2}}
(-8(G_{1} (2\omega \omega^{\ast} + 1) + \\
 \omega \omega^{\ast} (\omega \omega^{\ast} + 1) G'_{1}) n^{2} \bigg(4\lambda (1 + \omega \omega^{\ast})^{-4n} \bigg(-\frac{(\omega \omega^{\ast} + 1)^{4n} G_{1}}{2n(4\lambda)^{n}}\bigg)^{\frac{2}{2n-1}}\bigg)^{n} + \\
2^{2(n+1)} n \bigg(\lambda (1 + \omega \omega^{\ast})^{-4n} \bigg(-\frac{(\omega \omega^{\ast} + 1)^{4n} G_{1}(\omega\omega^{\ast})}{2n(4\lambda)^{n}}\bigg)^{\frac{2}{2n-1}}\bigg)^{n} G'_{1} + \\
 \bigg(-\frac{(\omega \omega^{\ast} + 1)^{4n} G_{1}}{2n (4\lambda)^{n}}\bigg)^{\frac{1}{2n-1}} ((((6n - 1) \omega \omega^{\ast} + 
2n - 1) G'_{1}(\omega \omega^{\ast}) + \\ 
2\omega \omega^{\ast} (1 + \omega \omega^{\ast}) (n - \frac{1}{2}) G''_{1}) G_{1} +
 \omega \omega^{\ast} (1 + \omega \omega^{\ast}) G'^{2}_{1}) (1 + \omega \omega^{\ast})) d \omega \bigg] - \\ 
2 \bigg(n - \frac{1}{2}\bigg) \bigg(-2^{2n+1}n\bigg(-\frac{(\omega \omega^{\ast} + 1)^{2} G_{1}}{2n}\bigg)^{\frac{2n}{2n-1}} + \\
G'_{1} \bigg(-\frac{(\omega \omega^{\ast} + 1)^{4n} G_{1}}{2n (4\lambda)^{n}}\bigg)^{\frac{1}{2n-1}} (1 + \omega \omega^{\ast}) \bigg) \omega \bigg\} d \omega^{\ast} \label{potential_kdef_ng}
\end{gathered}
\end{equation}

Hence, the Bogomolny equation for the k-deformed ungauged BPS baby Skyrme model, has the form:

 \begin{equation} 
 \varepsilon^{mn} i \ \omega_{,m}\omega^{\ast}_{,n} = \bigg(-\frac{(1+\omega\omega^{\ast})^{4n} 
  G_{1}(\omega \omega^{\ast})}{2n(4 \lambda)^{n}}\bigg)^{\frac{1}{2n-1}},
 \end{equation}

 if the potential has the form (\ref{potential_kdef_ng}). Some similar result was obtained in \cite{Bednarski2014}. 


\subsubsection{The gauged version}

  Now we apply the stereographic projection, for the functional (\ref{model1})
 
 \begin{equation}
 \vec{\phi} = \bigg[\frac{\omega + \omega^{\ast}}{1+\omega \omega^{\ast}}, \frac{-i \cdot (\omega-\omega^{\ast})}{1+\omega \omega^{\ast}}, \frac{1 - \omega 
 \omega^{\ast}}{1+\omega \omega^{\ast}}\bigg], \ \ i.e. \ \ \omega = \frac{\phi_{1}+i \phi_{2}}{1+\phi_{3}},
 \label{stereograf}
 \end{equation} 
 
 where $\omega=\omega(x,y) \in \mathbb{C}$, and $x, y \in \mathbb{R}$.\\ 

 We make the gauge transformation of $\mathcal{H} = \bigg[ 4 \lambda_{1} \frac{(i \varepsilon^{jk} \omega_{,j} \omega^{\ast}_{,k} - \varepsilon^{pr} A_{p} (\omega \omega^{\ast})_{,r})^{2}}{(1+\omega\omega^{\ast})^{4}} \bigg]^{n} + 
  V + \lambda_{2} U B^{2}$, on the sum of the invariants, and we have:

 \begin{gather}
 \mathcal{H} = \bigg[ 4 \lambda_{1} \frac{(i \varepsilon^{jk} \omega_{,j} \omega^{\ast}_{,k} - \varepsilon^{pr} A_{p} (\omega \omega^{\ast})_{,r})^{2}}{(1+\omega\omega^{\ast})^{4}} \bigg]^{n} + 
   V + \lambda_{2} U B^{2} + \\ 
  \lambda_{3} \{ G'_{1} [(i \varepsilon^{jk} \omega_{,j} \omega^{\ast}_{,k} - \varepsilon^{pr} A_{p} (\omega \omega^{\ast})_{,r})] + G_{1} B \} + \sum^{2}_{l=1} D_{l} G_{l+1}
 \end{gather}

 where obviously, $B \equiv F_{12} = \partial_{1} A_{2} - \partial_{2} A_{1}$ and $j, k, l, p, r = 1, 2$. We have assumed here again that $V =V(\omega, \omega^{\ast}, A_{k})$, $k = 1, 2$ (and $U=U(\omega, \omega^{\ast})$).

 So, the invariants are again: $ I_{1} = \lambda_{3} \{G'_{1} [(i\varepsilon^{ij}\omega_{,i}\omega^{\ast}_{,j}) - \varepsilon^{pr} A_{p} \cdot  (\omega \omega^{\ast})_{,r}] + G_{1} B \}$, ($\lambda_{3}=const$., $G_{1}=G_{1}(\omega \omega^{\ast}) \in \mathbb{R}$ is some arbitrary function differentiable at least twice, $G'_{1}$ denotes the derivative of the function $G_{1}$ with respect to its argument: $\omega\omega^{\ast}$), $I_{2}= D_{x} G_{2}(\omega, \omega^{\ast}, A_{1}, A_{2}), I_{3}=D_{y} G_{3}(\omega, \omega^{\ast}, A_{1}, A_{2})$. Next: $x_{1}=x, x_{2}=y, D_{x} \equiv \frac{d}{dx}, D_{y} \equiv \frac{d}{dy}$,  and $G_{l+1}=G_{l+1}(\omega, \omega^{\ast}, A_{1}, A_{2})$, ($l = 1, 2$), are some functions (differentiable at least twice), which are to be determinated later.
 
The dual equations have the form

 \begin{gather}
 \tilde{\mathcal{H}}_{,\omega} = \frac{1}{4} n \frac{4^{n} (\lambda_{1})^{n-1} N_{4}^{2(n-1)} }{(1+\omega \omega^{\ast})^{4n}}  
\bigg[ - \frac{8\lambda_{1} N_{4} \varepsilon^{pr} A_{p} \omega^{\ast}_{,r} }{(1+\omega\omega^{\ast})^{4}}  - 
\frac{16\lambda_{1} N^{2}_{4} \omega^{\ast}}{(1+\omega\omega^{\ast})^{5}}\bigg]  (1+\omega \omega^{\ast})^{4}  + \nonumber \\ 
V_{,\omega} + \lambda_{2} U_{,\omega} B^{2} + 
\lambda_{3} \{ G''_{1} \omega^{\ast} [i \varepsilon^{jk} \omega_{,j} \omega^{\ast}_{,k} - \varepsilon^{pr} A_{p} (\omega \omega^{\ast})_{,r}]  - \label{gornecechk1} \\  
G'_{1} \varepsilon^{pr} A_{p} \omega^{\ast}_{,r} + G'_{1} \omega^{\ast} B \} + \sum^{2}_{l=1} D_{l} G_{l+1, \omega} = 0, \nonumber  \\
\tilde{\mathcal{H}}_{,\omega^{\ast}} = \frac{1}{4} n \frac{4^{n} (\lambda_{1})^{n-1} N_{4}^{2(n-1)} }{(1+\omega \omega^{\ast})^{4n}}  
\bigg[ - \frac{8\lambda_{1} N_{4} \varepsilon^{pr} A_{p} \omega_{,r} }{(1+\omega\omega^{\ast})^{4}}  - 
\frac{16\lambda_{1} N^{2}_{4} \omega}{(1+\omega\omega^{\ast})^{5}}\bigg]  (1+\omega \omega^{\ast})^{4} +  \nonumber \\ 
 V_{,\omega^{\ast}} + \lambda_{2} U_{,\omega^{\ast}} B^{2} + 
\lambda_{3} \{ G''_{1} \omega [(i \varepsilon^{jk} \omega_{,j} \omega^{\ast}_{,k} - \varepsilon^{pr} A_{p} (\omega \omega^{\ast})_{,r})] - \label{gornecechk2} \\  
G'_{1} \varepsilon^{pr} A_{p} \omega_{,r} + G'_{1} \omega B \} + \sum^{2}_{l=1} D_{l} G_{l+1, \omega^{\ast}} = 0, \nonumber
\end{gather}

  \begin{gather} 
  \tilde{\mathcal{H}}_{,A_{s}}  : \   -2n \frac{(4)^{n} \lambda^{n}_{1} N^{2n-1}_{4}}{(1+\omega \omega^{\ast})^{4n}} (\varepsilon^{sr}(\omega \omega^{\ast})_{,r})  + V_{,A_{s}} - \nonumber \\ 
  \lambda_{3}G'_{1}(\varepsilon^{sr} (\omega \omega^{\ast})_{,r}) + \sum^{2}_{l=1}  D_{l}G_{l+1,A_{s}} = 0,  \label{gornecechk3} \\
  \tilde{\mathcal{H}}_{,\omega_{s}} = 2n \frac{(4)^{n} \lambda^{n}_{1} N^{2n-1}_{4}}{(1+\omega \omega^{\ast})^{4n}} (i\varepsilon^{sk} \omega^{\ast}_{,k} - \varepsilon^{ps} A_{p} \omega^{\ast} )  + \\
  \lambda_{3} G'_{1} (i \varepsilon^{sk} \omega^{\ast}_{,k} - \varepsilon^{ps} A_{p} \omega^{\ast}) + G_{s+1, \omega} = 0, \nonumber
 \end{gather}

\begin{gather}
\tilde{\mathcal{H}}_{,\omega^{\ast}_{s}} = 2n \frac{(4)^{n} \lambda^{n}_{1} N^{2n-1}_{4}}{(1+\omega \omega^{\ast})^{4n}} (i\varepsilon^{js} \omega_{,j} - \varepsilon^{ps} A_{p} \omega )  +  \\ 
\lambda_{3} G'_{1} (i \varepsilon^{js} \omega_{,j} - \varepsilon^{ps} A_{p} \omega) + G_{s+1, \omega^{\ast}} = 0, \nonumber
\end{gather}

 \begin{gather} 
 \tilde{\mathcal{H}}_{,A_{s,r}}  : 2 \lambda_{2} U B \varepsilon^{rs}+ G_{1} \varepsilon^{rs} + G_{r+1,A_{s}} = 0
 \end{gather}

 where $N_{4} = [i \cdot (\varepsilon^{jk} \omega_{,j}\omega^{\ast}_{,k}) - \varepsilon^{pr} A_{p} \cdot  (\omega \omega^{\ast})_{,r}]$ and $G'_{1}, G''_{1}$ denote the derivatives of the function $G_{1}$ with respect to its argument: $\omega\omega^{\ast}$.

 Now, in order to make the system self-consistent, we put

 \begin{gather}
 G_{2} = c_{2} A_{2}, \ G_{3} = - c_{2} A_{1}, \\
 i \varepsilon^{jk} \omega_{,j} \omega^{\ast}_{,k} - \varepsilon^{pr} A_{p} (\omega \omega^{\ast})_{,r} = \bigg( - \frac{\lambda_{3} G'_{1}}{2n(4\lambda_{1})^{n}} (1 + \omega + \omega^{\ast})^{4n} \bigg)^{\frac{1}{2n-1}},\\
 B = - \frac{\lambda_{3} G_{1} + c_{2}}{2 \lambda_{2} U}
 \end{gather}

 and we insert these relations, into (\ref{gornecechk1}) - (\ref{gornecechk3}). Hence, we get the equations for $V$ and $U$. We get the formula for $V$ (if $G_{1}=G_{1}(\omega \omega^{\ast})$):

\begin{equation}
\begin{gathered}
 V(\omega, \omega^{\ast}) =  \frac{1}{ \lambda_{2} U} \bigg\{ \int \frac{1}{(2n-1) (\omega \omega^{\ast} + 1)} \bigg\{ (\omega \omega^{\ast} + 1)   \\ 
\bigg[ \int \frac{1}{G'_{1} (\omega \omega^{\ast} + 1)^{2}} (-8(G'_{1} (2\omega \omega^{\ast} + 1) + 
 \omega \omega^{\ast} (\omega \omega^{\ast} + 1) G''_{1}) n^{2}   \\ 
\bigg( 4 \lambda_{1} (\omega\omega^{\ast} + 1)^{-4n} \bigg(-\frac{\lambda_{3}(\omega \omega^{\ast} + 1)^{4n} G'_{1}}{2n(4\lambda)^{n}}\bigg)^{\frac{2}{2n-1}}\bigg)^{n} + \\
2^{2(n+1)} n \bigg(\lambda_{1} (1 + \omega \omega^{\ast})^{-4n} \bigg(- \lambda_{3}\frac{(\omega \omega^{\ast} + 1)^{4n} G'_{1}}{2n(4\lambda_{1})^{n}}\bigg)^{\frac{2}{2n-1}}\bigg)^{n} G'_{1} +   \\
 2 \lambda_{3} \bigg(-\lambda_{3} \frac{(\omega \omega^{\ast} + 1)^{4n} G'_{1}}{2n (4\lambda)^{n}}\bigg)^{\frac{1}{2n-1}} \bigg( \bigg( \bigg( \bigg( 3n - \frac{1}{2} \bigg) \omega \omega^{\ast} + \\ 
n - \frac{1}{2} \bigg) G''_{1} + \omega \omega^{\ast} (1 + \omega \omega^{\ast}) \bigg(n - \frac{1}{2}\bigg) G'''_{1} \bigg) G'_{1} +  \\ 
 \frac{1}{2} \omega \omega^{\ast} (1 + \omega \omega^{\ast})(G''_{1})^{2}\bigg) (1 + \omega \omega^{\ast})) d \omega \bigg] -  \\ 
2\bigg(-2^{2n+1}n\bigg(\lambda_{1} (1+\omega\omega^{\ast})^{-4n} \bigg(-\frac{(\omega \omega^{\ast} + 1)^{4n} G'_{1}}{2n (4\lambda_{1})^{n}}\bigg)\bigg)^{\frac{2n}{2n-1}} +  \\
 \lambda_{3} G''_{1} \bigg(-\frac{-\lambda_{3}(\omega \omega^{\ast} + 1)^{4n} G'_{1}}{2n (4\lambda_{1})^{n}}\bigg)^{\frac{1}{2n-1}} (1 + \omega \omega^{\ast}) \bigg) \omega \bigg(n-\frac{1}{2} \bigg) \bigg\} d \omega^{\ast} \lambda_{2} U +  \\
4\lambda_{2}U \int \bigg[ \bigg( -2^{2(n+1)} \bigg( \bigg(-\lambda_{1} (1+\omega \omega^{\ast})^{-4n} \frac{\lambda_{3}(\omega \omega^{\ast} + 1)^{4n} G'_{1}}{2n(4\lambda)^{n}}\bigg)^{\frac{2}{2n-1}} \bigg)^{n} n + \\
G''_{2} \bigg(-\frac{\lambda_{3}(\omega \omega^{\ast} + 1)^{4n} G'_{1}}{2n(4\lambda)^{n}}\bigg)^{\frac{1}{2n-1}} \lambda_{3} (1+\omega \omega^{\ast}) \bigg) \omega^{\ast} (1 + \omega \omega^{\ast})^{-1} \bigg] d \omega  +  \\
+ 4 c_{3} \lambda_{2} U + (\lambda_{3} G_{1} + c_{2})^{2} \bigg\}. \label{potential_kdef_ng}
\end{gathered}
\end{equation}

\section{Summary}
In the present paper BPS equations for some generalization of the gauged baby BPS Skyrme model, have been found. This have been performed by applying the concept of strong necessary conditions (CSNC).

In the case of the non-minimally coupled gauged baby BPS Skyrme model, we found the the new BPS equation is modified by a coupling between the magnetic field $B$ and the dielectric function $U$, in both cases: for an axially symmetric ansatz and for the energy functional expressed by stereographic variables. 
In the case of the ansatz, the term (in BPS equation), which emerges due to the gauge coupling (proportional to $g^2$) is modified by $1/U$. This modification can lead to some new restriction on possible potentials $V$ (and the dielectric functions $U$), for which the BPS equation has nontrivial topological solutions.  $U$ does not depend on the field $a_{2}$ (where $A_{\phi}=ka_{2}(r)$), but $V$ depends on it in the general case. Hence, we have an analogon to Proca theory. However, in the case of (\ref{z_U2}), if the functions $f_{1} = f_{2}$, then $V=V(h)$.  
Another modifications of the Bogomolny decomposition and the formula for the potential, can be observed, if one compares these results (for the case with stereographic variables), with the results obtained for gauged restricted baby BPS Skyrme model with minimal coupling,  \cite{stepien2015}. 
\\
For k-deformation (given by polynomial function $G_{0}$), both the Bogomolny equation for the matter (Skyrme) field, as well as the superpotential equation, are modified. \\
In all these cases of gauged baby BPS Skyrme model, investigated in this paper, expressed in stereographic variables, the potential (for which Bogomolny decomposition exists), does not depend on the gauge field $A_{k} \ (k =1, 2)$, so in general situation, we have not the Proca theory (cf. \cite{stepien2015}).

As the superpotential equation leads to some nontrivial restrictions on the Skyrme potential $V$, for which solutions of the Bogomolny equations exist, it would be desirable to study it in detail. 

Another direction is to investigate a relation between the CSNC construction and supersymmetry (which always is hidden behind 
Bogomolny equations \cite{sokalski2002}, \cite{susy}). 

In any case, the CSNC framework proven to be a powerful and strightforward method for derivation of the Bogomolny equations. 

\section{Acknoweledgements}
  
  The author thanks to Dr. Hab. A. Wereszczynski for interesting discussions. \\

  \section{Computational resources}
  
     This research was supported in part by PL-Grid Infrastructure.

\end{document}